\documentclass[aps,prb,twocolumn,showpacs]{revtex4}
\usepackage{graphicx}
\usepackage{amsmath}
\usepackage{float}

\usepackage{appendix}
\usepackage{subfigure}
\usepackage{times}
\usepackage{natbib}

\newcommand{\beq}{\begin{equation}}
\newcommand{\eeq}{\end{equation}}
\newcommand {\ba} {\begin{eqnarray}}
\newcommand {\ea} {\end{eqnarray}}

\begin{document}

\title{Optical properties of the optimally doped Ca$_{8.5}$La$_{1.5}$(Pt$_3$As$_8$)(Fe$_{2}$As$_{2}$)$_5$ single crystal}

\author{Yu-il Seo$^1$}
\author{Woo-Jae Choi$^1$}
\author{Shin-ichi Kimura$^2$}
\author{Yunkyu Bang$^{3}$}
\email[]{ykbang@chonnam.ac.kr}
\author{Yong Seung Kwon$^1$}
\email[]{yskwon@dgist.ac.kr}

\affiliation{$^1$Department of Emerging Materials Science, Daegu Gyeongbuk
Institute of Science and Technology (DGIST), Daegu 711-873, Republic of Korea \\
$^2$FBS and Department of Physics, Osaka University, Suita 565-0871, Japan \\
$^3$Department of Physics, Chonnam National University, Kwangju
500-757, Republic of Korea}

\begin{abstract}
We have measured the reflectivity of the optimally doped Ca$_{8.5}$La$_{1.5}$(Pt$_3$As$_8$)(Fe$_{10}$As$_{10}$) single crystal ($T_c$ = 32.8K) over the broad frequency range from 40 $cm^{-1}$ to 12000 $cm^{-1}$ and for temperatures from 8K to 300 K.
The optical conductivity spectra of the low frequency region ($< 1,000$ $cm^{-1}$) in the normal state (80 K $< T \leq$ 300 K) is well fitted with two Drude forms, which indicates the presence of multiple bands at the Fermi level. Decreasing temperature below 80 K,
this low frequency Drude spectra develops pseudogap (PG) hump structure at around $\approx 100$ $cm^{-1}$ and continuously evolves into the fully opened superconducting (SC) gap structure below $T_c$.
Theoretical calculations of the optical conductivity with the preformed Cooper pair model provide an excellent description of the temperature evolution of the PG structure above $T_c$ into the SC gap structure below $T_c$.
The extracted two SC gap sizes are $\Delta_S$ = 4.9 $meV$  and $\Delta_L$ = 14.2 $meV$, suggesting  Ca$_{8.5}$La$_{1.5}$(Pt$_3$As$_8$)(Fe$_{10}$As$_{10}$) as a multiple gap superconductor with a mixed character of the weak coupling and strong coupling superconductivity.
\end{abstract}


\date{\today}
\maketitle

\section{Introduction}
Recently discovered Fe-pnictide compounds, Ca$_{10}$(Pt$_3$As$_8$)(Fe$_{2}$As$_{2}$)$_5$ (so-called 10-3-8 compound) and Ca$_{10}$(Pt$_4$As$_8$)(Fe$_{2}$As$_{2}$)$_5$
(so-called 10-4-8 compound) \cite{ni2011,kakiya2011,sturzer2012,sapkota2014}, added another new class to the family of the Fe-based superconductors (IBSs).
They are showing the prototype behavior of the subtle balance and competition between magnetism and superconductivity commonly observed in other IBSs. Both compounds share similar overall crystal structure, consisting of tetrahedral FeAs planes sandwiched between the planar Pt$_n$As$_8$ ($n=3, 4$) intermediary layers. However, in fine details, there exist also distinct differences between them.

The parent 10-4-8 compound has a tetragonal crystal structure and is metallic. It becomes superconducting (SC) without doping with the maximum $T_c$ of $\sim 38$ K \cite{ni2011,johrendt2011}, and doping of electrons only suppresses the SC transition temperature\cite{struzer2013,sapkota2014}.
On the other hand, the parent 10-3-8 compound has a triclinic crystal structure. Sturzer {\it et al.} \cite{struzer2013} reported that this compound is semiconducting and becomes antiferromagnetically ordered below $T_N \approx 120$ K without any further reduction in the crystal symmetry. Superconductivity can be induced by doping of electrons: for instance, La substitution at the Ca site, or Pt substitution at the Fe site. The maximum $T_c$ for each substitution can be $\sim 32$ K and $\sim 15$ K, respectively \cite{struzer2013}. However, Neupane {\it et al.} \cite{neupane2012} reported that the undoped 10-3-8 compound has the multiple Fermi surfaces like other typical FeAs-superconducting compounds and becomes superconducting at $T_c \sim 8K$ without doping.

The band structure calculations\cite{shein2011} suggested that these differences arise from the increased metallicity of the PtAs layers.
And recent angle resolved photoemission spectroscopy (ARPES) experiment further suggested that this difference between two compounds could be developed by the number of the band-edge singularities \cite{borisenko2013}.
In order to  study this subtle relation between the electronic structure and superconductivity in these materials, various experiments have already been performed:
such as transport \cite{ni2013}, pressure effect \cite{gao2014}, ARPES \cite{neupane2012}, magnetic force microscope (MFM) \cite{movshovich2012},
upper critical field ($H_{c2}$) \cite{mun2012}, nuclear magnetic resonance (NMR) \cite{zhou2013}, and penetration depth $\lambda(T)$\cite{prozorov2014} measurements, etc.
However, the IR spectroscopy experiment has not yet been carried out with these compounds.
This technique is particularly a useful tool to directly study the low energy dynamics of the correlated materials with temperature variation, and, therefore, it is able to investigate the systematic development of superconductivity with temperature and any non-trivial evolution of the correlation effects, if exists.

In this paper, we have measured the IR reflectivity and analyzed the optical properties of the optimally La-doped 10-3-8 compound, Ca$_{8.5}$La$_{1.5}$(Pt$_3$As$_8$)(Fe$_{2}$As$_{2}$)$_5$ single crystal ($T_c \approx $32.8 K).
In view of the phase diagram with electron doping of the previous study\cite{ni2013}, our compound is located far from the antiferromagnetic (AFM) quantum critical point (QCP) where $T_N$ goes to zero in the phase diagram. However, we speculate that our compound is located near another non-magnetic QCP in the universal phase diagram, from the observations of (1) the maximum $T_c$ with doping variation, and (2) the $T$-linear resistivity data above $T_c$ up to 300 K.
It is interesting to note that these QCP behaviors far away from the AFM QCP are similar to the case of the cuprate superconductors.
Then, most interestingly, we have also found the pseudogap (PG) behavior in the optical conductivity of our sample up to the temperature about three times higher ($\sim 80$ K) than $T_c (\approx$ 33 K). With the observation that our compound is far away from the AFM QCP, it is logical to conclude that the PG behavior of the optimal doped 10-3-8 compound is not related to the AFM correlation. We have employed the preformed Cooper pair model to calculate the real part of the optical conductivity $\sigma(\omega, T)$ and successfully reproduced the PG feature far above $T_c$ and its consistent evolution to the SC gap structure below $T_c$.

Combining the previous reports of the PG observations in several IBS compounds and cuprate superconductors, our observation establishes that the PG phenomena induced by a SC correlation is a generic phenomena of the strongly correlated unconventional superconductors.
\section{Experimental Details}
\begin{figure}
\includegraphics[width=100mm]{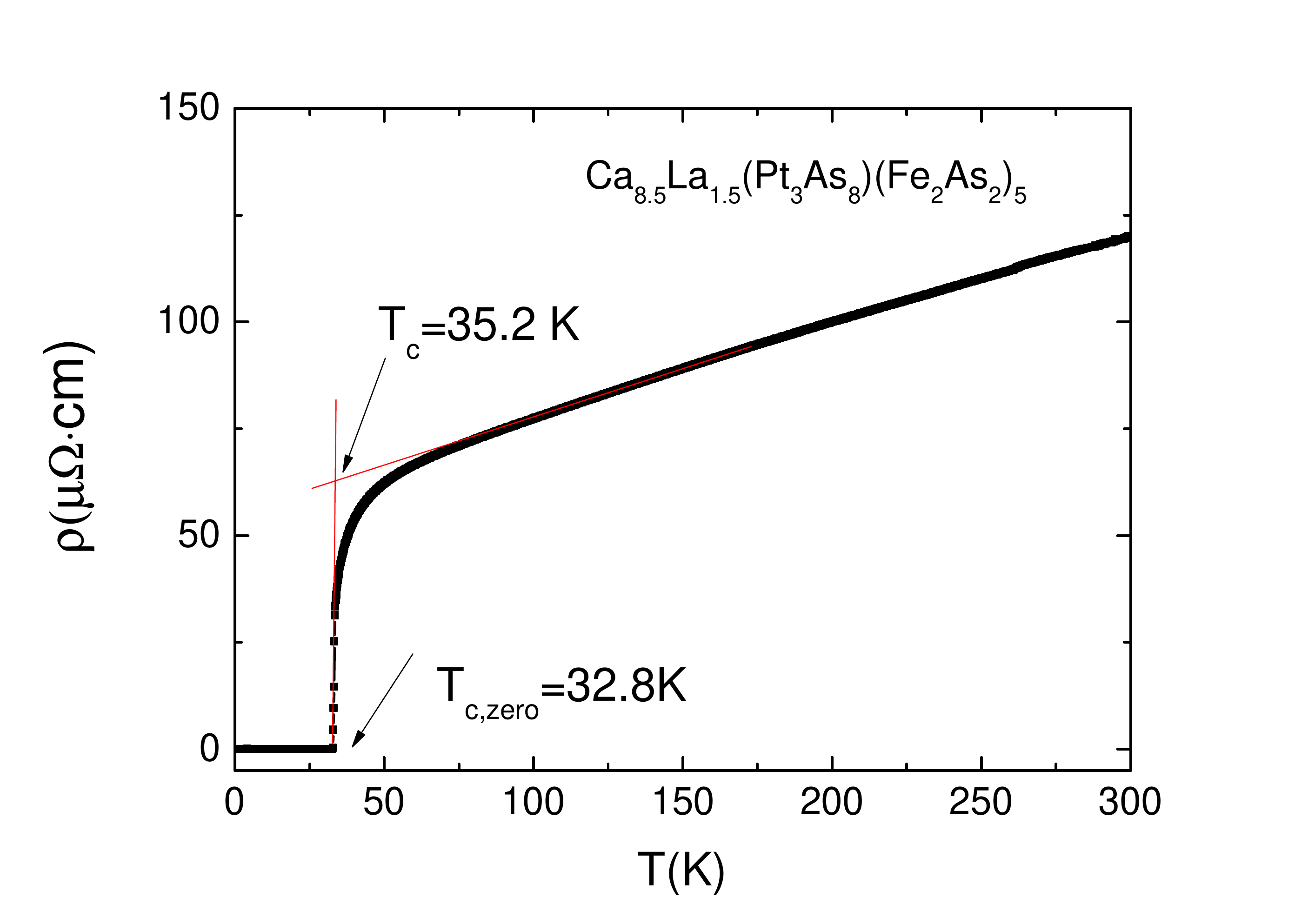}
\vspace{-0.cm}
\caption{(Color online) Temperature dependence of the DC electrical resistivity $\rho(T)$ of the Ca$_{8.5}$La$_{1.5}$(Pt$_3$As$_8$)(Fe$_{2}$As$_{2}$)$_5$ single crystal. \label{fig1}}
\end{figure}

High quality single crystals of Ca$_{8.5}$La$_{1.5}$(Pt$_3$As$_8$)(Fe$_{2}$As$_{2}$)$_5$ were grown by a Bridgeman method with a sealed molybdenum crucible at 1250$^{\circ}$ C.
Before performing the Bridgeman method, we made the precursors of CaAs, LaAs and FeAs in advance.
Figure 1 shows the in-plane electrical resistivity $\rho(T)$ of the Ca$_{8.5}$La$_{1.5}$(Pt$_3$As$_8$)(Fe$_{2}$As$_{2}$)$_5$  single crystal.
We do not see any noticeable anomaly in $\rho(T)$ around $T_N^0 \approx 120$ K  (the Neel temperature of the undoped parent 10-3-8 compound) and below due to the spin density wave (SDW) transition.
The SC transition begins at $T_c$=35.2 K and zero resistivity occurs at $T_{c,zero}=32.8$ K, respectively. These transition temperatures are about 7-9 K higher than the one with similar nominal doping of La ($x=1.45$) in the previously reported work \cite{ni2013}, indicating that our sample is a substantially higher quality single crystal.

Another noticeable feature is that $\rho(T)$ shows the $T$-linear temperature dependence above $T_c$ up to 300 K. This linear-in-$T$ behavior in $\rho(T)$ is commonly observed in other optimally doped IBSs \cite{kamihara2008,chen2008,ren2008,hsu2008} and is a signature that the sample is near the QCP. This is consistent with the phase diagram of Ref.\cite{ni2013}, where the maximum SC $T_c$ was also located with the La doping "$x_{max}^{sc}$" $\approx1.5$. Moreover the phase diagram of Ref.\cite{ni2013} shows that the magnetic critical doping "$x_c^{mag}$", where $T_N \rightarrow 0$,  is near $\approx 0.3$, which is far apart from the optimal doping $x_{max}^{sc}$ for the maximum SC $T_c$. Therefore, in view of the phase diagram of Ref.\cite{ni2013} and our result, we speculate that there exist two QCPs in the La-doped 10-8-3 compound: one is the QCP due to magnetic fluctuations where $T_N \rightarrow 0$, and the other is the QCP due to unknown quantum fluctuations where the SC $T_c$ becomes maximum. The former type of QCP is more common with many IBSs as well as most of heavy fermion superconductors, while the latter type of QCP is well represented in the high-$T_c$ cuprate superconductors.
Therefore, clarifying the nature of the QCP in the 10-3-8 compound will be particularly interesting in connection with the mystery of the high-$T_c$ cuprate superconductors.

In order to investigate the electronic structure as well as  the SC properties of Ca$_{8.5}$La$_{1.5}$(Pt$_3$As$_8$)(Fe$_{2}$As$_{2}$)$_5$, we have measured the optical reflectivity $R(\omega,T)$ of this single crystal in a broad frequency range from 40 to 12000 $cm^{-1}$ and for various temperatures from 8 to 300 K. We used a Michelson-type rapid-scan Fourier spectrometer (Jasco FTIR610). In particular, we used a specially designed feedback positioning system to drastically reduce the overall uncertainty level. The uncertainty of our data was maintained less than 0.5$\%$.
For more details of experimental methods of reference setting and control of data uncertainty, we refer to our previous paper\cite{kwon2012}.

\section{Results and Discussions}
\subsection{Reflectivity and Optical Conductivity}
Figure 2 shows the reflectivity spectra $R(\omega,T)$ of Ca$_{8.5}$La$_{1.5}$(Pt$_3$As$_8$)(Fe$_{2}$As$_{2}$)$_5$ single crystal at several different temperatures
from 8 K to 300 K. The main panel shows the full measured data for frequencies from 50 $cm^{-1}$ to 12000 $cm^{-1}$ in log scale, and the inset shows the close-up view of the data from 0 $cm^{-1}$  to 300 $cm^{-1}$ in linear scale; here the data from 0 $cm^{-1}$ to 60 $cm^{-1}$ (dotted lines) are not the measured ones but extrapolated ones as explained below (although we have data from 40 $cm^{-1}$, we didn't use the data of 40 $cm^{-1}$ - 60 $cm^{-1}$ because the data in this part are not uniformly clean).

In the normal state (the data sets for $T \geq 40$ K), $R(\omega)$ approaches to unity at zero frequency and increases as decreasing temperature in far-infrared region, showing a typical metallic behavior of Ca$_{8.5}$La$_{1.5}$(Pt$_3$As$_8$)(Fe$_{2}$As$_{2}$)$_5$.
Upon entering the SC phase (the data sets below 40 K), the low frequency reflectivity turns up quickly, and reaches the flat unity response below $\sim $80 $cm^{-1}$ reflecting the SC gap opening. As the temperature increases toward $T_c$ the flat unity response shrinks to the lower frequency region as the SC gap size is decreasing. The shape of the flat part in $R(\omega)$ at low frequencies and its temperature dependence are clear signatures of the fully opened SC gap. However, the most interesting development of the reflectivity spectra in Fig.2 occurs in the data sets for 80 K (cyan color) and 40 K (purple color), which are still in the normal state. Both data sets show a curious suppression of spectra in the low frequency region around 100 $cm^{-1}$ (see the inset of Fig.2).

For more convenient analysis, we converted our reflectivity data $R(\omega)$ into the real part of the optical conductivity $\sigma_1(\omega)$ by the Kramers-Kronig (KK) transformation. For extrapolation of the data for the KK-transformation, at low frequencies, we used the Hagen-Rubens extrapolation formula for the normal state and the form ($1-A\omega^4$) below the gap in the SC state. For high frequencies above 12000 $cm^{-1}$, the standard  $1/\omega^4$ form plus a constant reflectivity are used up to 40 $eV$.

\begin{figure}[h]
\noindent
\includegraphics[width=160mm]{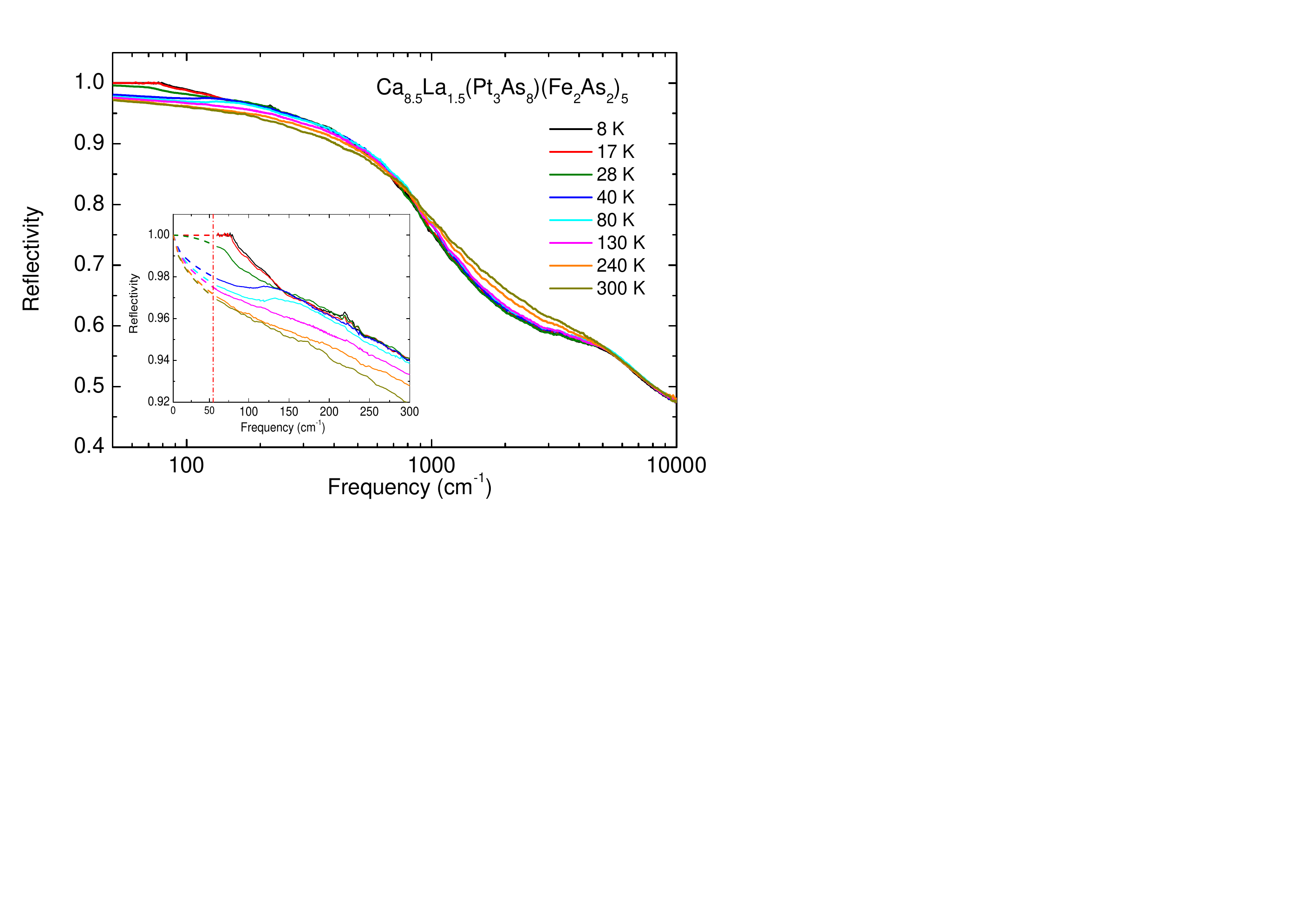}
\vspace{-5cm}
\caption{(Color online) Reflectivity spectra $R(\omega)$ of the Ca$_{8.5}$La$_{1.5}$(Pt$_3$As$_8$)(Fe$_{2}$As$_{2}$)$_5$ single crystal at several temperatures. The inset shows the enlarged view of spectra below 300 $cm^{-1}$; dotted lines are the part from extrapolation as explained in the main text. \label{fig2}}
\end{figure}
\begin{figure}[h]
\hspace{-1.2cm}
\includegraphics[width=100mm]{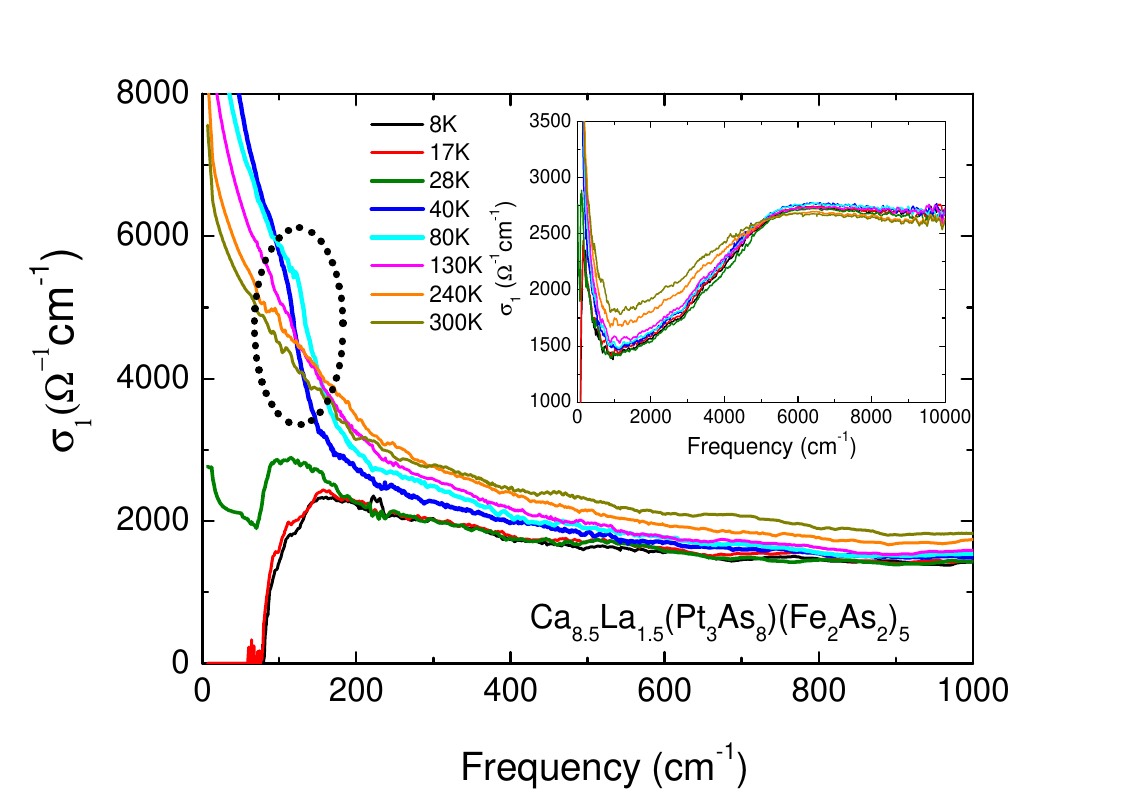}
\vspace{-0.5cm}
\caption{(Color online) Real part of optical conductivity $\sigma_1 (\omega)$ of Ca$_{8.5}$La$_{1.5}$(Pt$_3$As$_8$)(Fe$_{2}$As$_{2}$)$_5$ single crystal for 8, 17, 28, 40, 80, 130, 240, and 300 K, respectively. The pseudogap feature of the 40K and 80 K data is indicated inside the dotted circle.
The inset shows the same data $\sigma_1 (\omega)$ in wider view up to 10,000 $cm^{-1}$. \label{fig3}}
\end{figure}

Figure 3 shows these results of $\sigma_1(\omega)$, obtained from the data of Fig.2, in the frequency range from 0 to 1000 $cm^{-1}$ for various temperatures.
The optical conductivity spectra at 130, 240 and 300 K --  which are in normal state -- show a very broad Drude peak centered at $\omega =0$ and then monotonically decrease until the interband transition starts appearing around 1000 $cm^{-1}$ (see the inset of Fig.3). The interband transition spectra continues to form a broad hump around 6000 $cm^{-1}$; while the peak position of the hump is almost temperature independent, there exists an interesting spectral weight transfer with temperature variation which will be analyzed in section III.B.

As the temperature decreases down to 80 K (cyan color) and 40 K (blue color) -- which are still in normal state by transport/thermodynamic measurements -- (1) the Drude part of spectra rapidly sharpens, and then (2) a peculiar hump structure around 80-150 $cm^{-1}$ appears on top of the smooth Drude spectra.
The former is the expected evolution due to the formation of coherent quasiparticles as temperature decreases.
The striking feature is the latter, namely, the appearance of the shoulder-like hump structure far above $T_c$ on top of the Drude response in the frequency range of 80-150 $cm^{-1}$, indicating an incomplete gap -- hence called PG -- formation over the Fermi surfaces or in a part of the Fermi surfaces.
Then below $T_c$, i.e., at 28 K, 17 K and 8 K, $\sigma_1(\omega)$ shows a clean opening of the SC energy gap with the absorption edge at about 80 $cm^{-1}$ and a peak position at around 150 $cm^{-1}$. Comparing this SC gap structure at around  80-150 $cm^{-1}$  below $T_c$, the hump structure around 80-150 $cm^{-1}$, above $T_c$ at 40 K and 80 K, appears to be a continuous evolution of the SC correlation below $T_c$.

The observation of the PG features in the IBSs with the optical spectroscopy measurements is now quite common.
For example, the underdoped ($x\approx 0.12, 0.2$) \cite{dai2012} and the slightly underdoped ($x\approx 0.3$) \cite{kwon2012} Ba$_{1-x}$K$_x$Fe$_2$As$_2$ have reported the similar behavior in their optical measurements: namely, the hump structures in the optical conductivity  $\sigma_1(\omega)$ above $T_c$ (about three times of $T_c$ for all cases) in the same energy scale as the SC gap energy, and their continuous evolution into the SC gap. These authors concluded that this hump structure is not related with a magnetic correlation but rather connected to the SC gap, hence possibly a precursor of the preformed Cooper pairs. The optimally Co-doped Ba(Fe$_{0.92}$Co$_{0.08}$)As$_2$\cite{lobo2010} also has shown a similar hump structure at 30 K ($T_c$=22.5 K) and its evolution into the SC gap below $T_c$; however, the authors of Ref.\cite{lobo2010} advocated, as its origin, the impurity bound state or a low-energy interband transition, with which we do not agree.

More recently, Surmach et al. \cite{surmach_prb_2015} have performed a comprehensive measurements of muon-spin relaxation ($\mu$SR), inelastic neutron scattering (INS), and NMR on Pt-doped (CaFe$_{1-x}$Pt$_{x}$As)$_{10}$Pt$_{3}$As$_{8}$ ($T_c$ = 13 K) -- the same 10-3-8 compound we studied in this paper but with Pt doping on Fe sites. They found a PG behavior in the $1/T_1$ data of $^{75}$As NMR below $T^{\ast} \approx 45$ K (about three times higher than $T_c$). These authors concluded from a combination of measurements of $\mu$SR, INS, together with NMR  that this PG behavior is likely to be associated with the preformed Cooper pairs.
Therefore, our observation of a similar PG behavior in the optimally doped Ca$_{8.5}$La$_{1.5}$(Pt$_3$As$_8$)(Fe$_{2}$As$_{2}$)$_5$ compound has strengthened the universal nature of the PG phenomena originating from the preformed Cooper pairs or a precursor effect from the SC correlation in the 10-3-8 compound.

\begin{figure}
\noindent
\includegraphics[width=95mm]{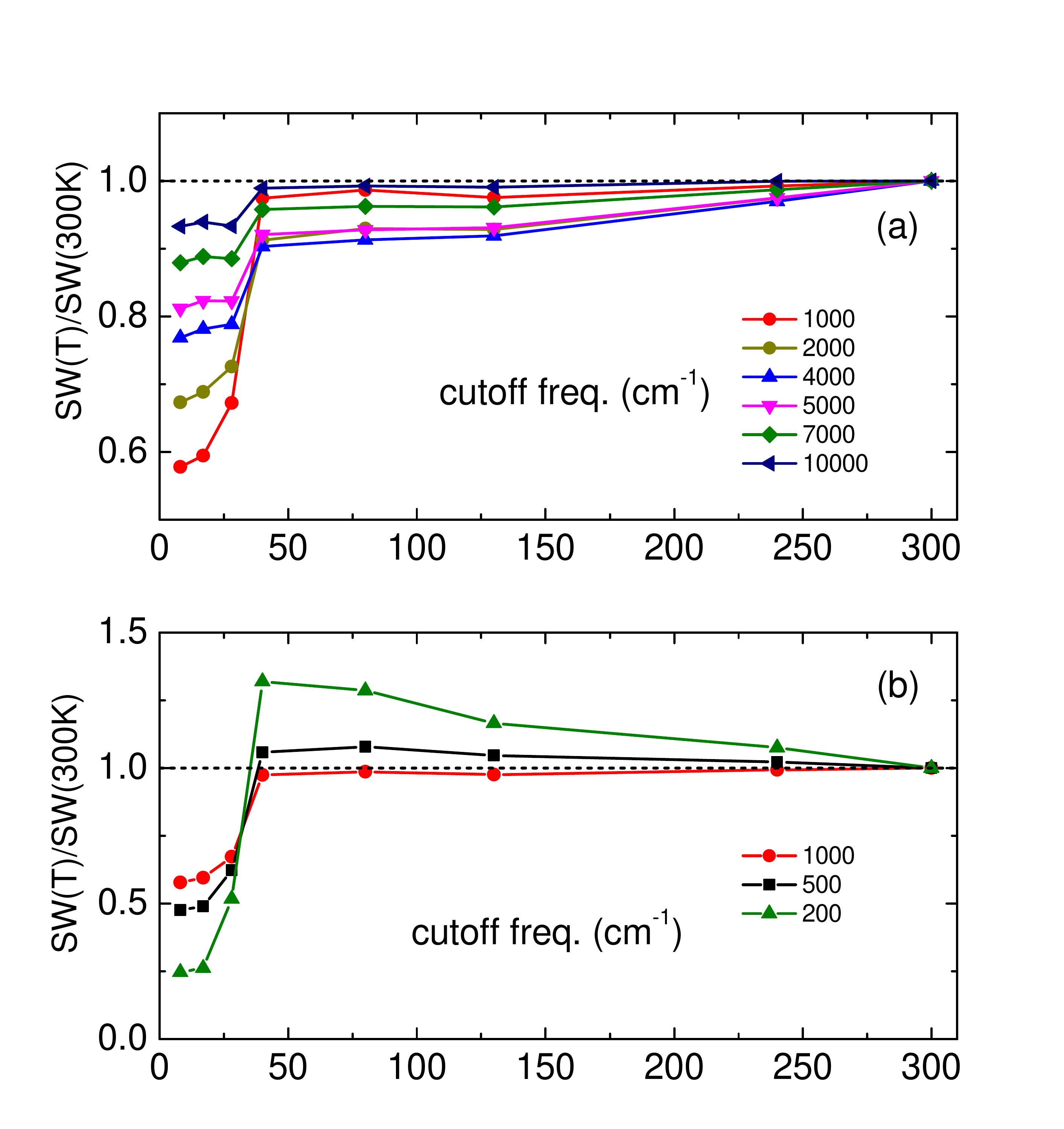}
\includegraphics[width=95mm]{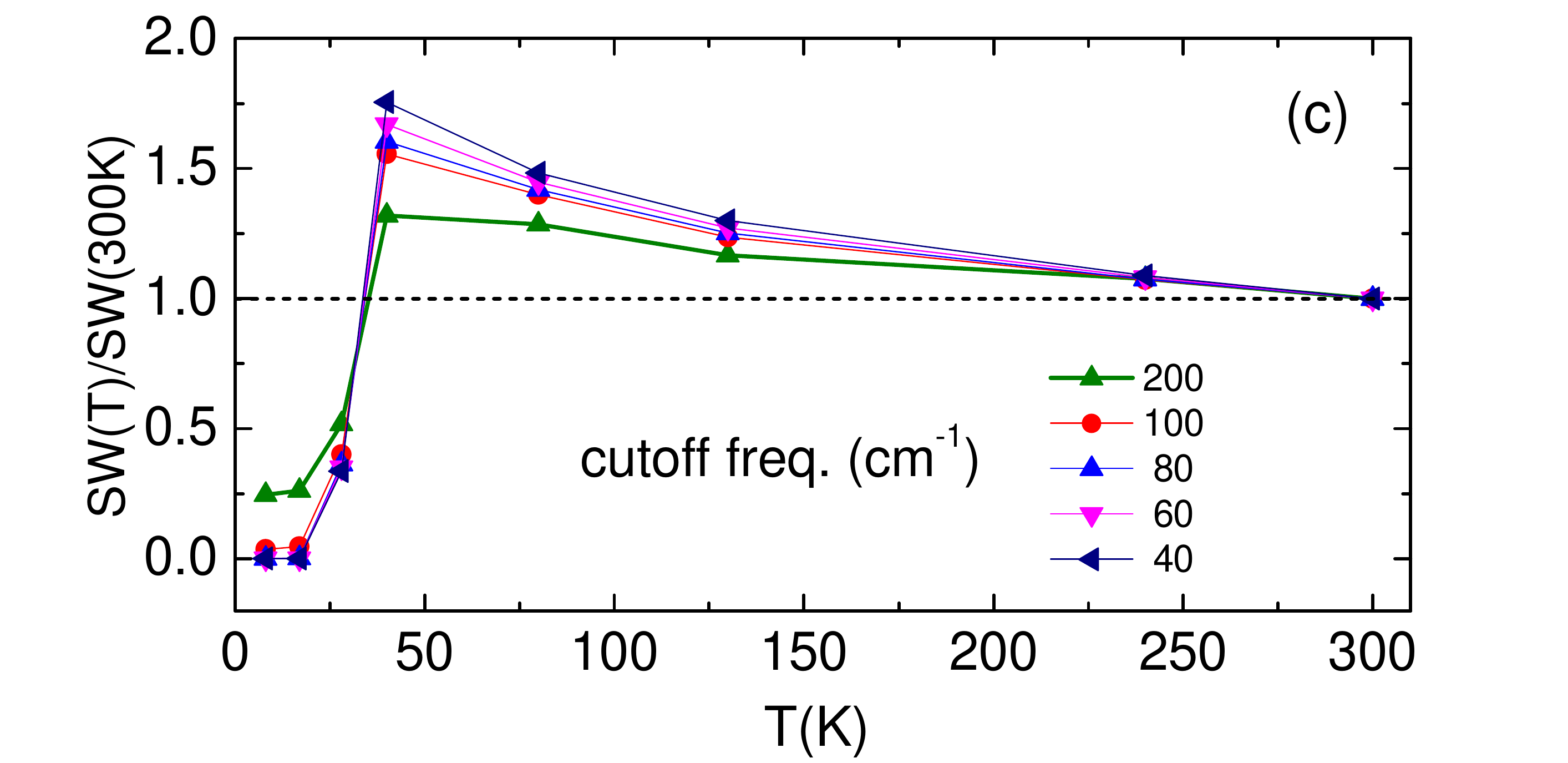}
\vspace{-.5cm}
\caption{(Color online) Temperature dependence of the normalized "partial" spectral weights $SW(T; \omega_c)/SW(300$ $K; \omega_c)$ of the optical conductivity data in Fig.3 with different cutoff frequency $\omega_c$. (A) data with $ \omega_c \geq 1000 cm^{-1}$; (B) data with $\omega_c \leq 1000 cm^{-1}$; and (C) data with finer variation of cutoff frequencies for $\omega_c \leq 200  cm^{-1}$: 200, 100, 80, 60, and 40 $cm^{-1}$, respectively.
\label{fig4}}
\end{figure}

\begin{figure}
\includegraphics[width=100mm]{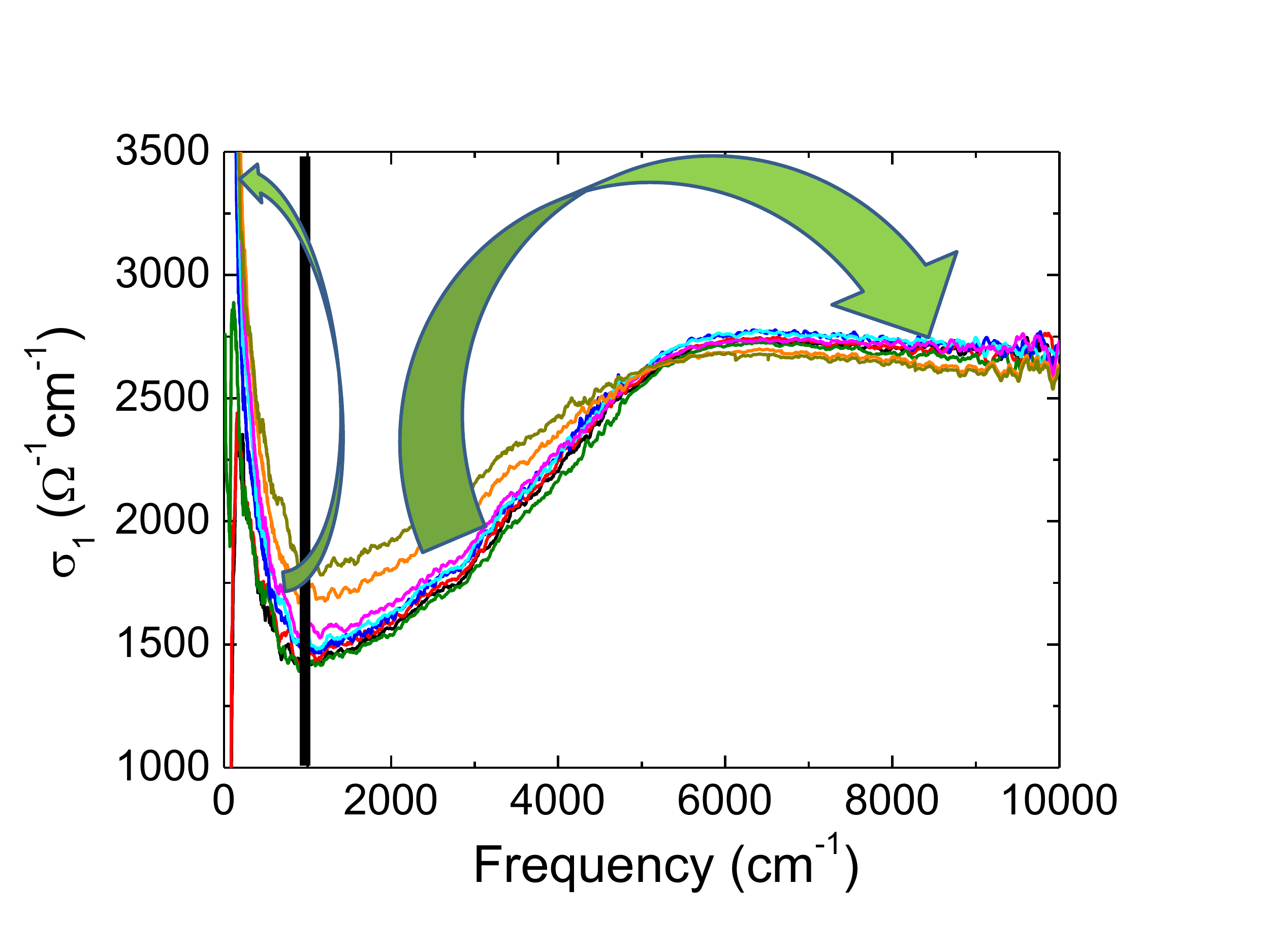}
\vspace{-0.cm}
\caption{(Color online) A schematic illustration of the spectral weight transfer with temperature variation. The spectral density in mid-frequency range ($\sim 500$ $cm^{-1} <  \omega < \sim 5,000$ $cm^{-1}$) is continuously depleted with decreasing temperature. This depleted spectral weight is roughly divided by $\omega^{\ast} \sim 1,000$ $cm^{-1}$ (black vertical line), and the spectral weight below $\omega^{\ast}$ and the spectral weight above $\omega^{\ast}$ are separately conserved (of course, this separate conservation rule holds only approximately).
\label{fig5}}
\end{figure}

\subsection{Temperature Dependence of Spectral Weight Transfer}
To further understand the temperature evolution of the optical conductivity shown in Fig.3, we have analyzed the temperature dependence of the "partial" spectral weight $SW(T; \omega_c)$ with different cutoffs defined as

\begin{equation}
SW(T; \omega_c) = \int^{\omega_c}_{0+} \sigma_1(\omega;T) d \omega,
\end{equation}
\noindent
where $\omega_c$ is a cutoff frequency.
Figure 4 shows the results of $SW(T; \omega_c)$, normalized by $SW(T=300K; \omega_c)$, that reveal non-trivial information of the correlated electron system.

Fig.4(A) shows that the total spectral weight up to 10,000 $cm^{-1}$ is constant on changing temperature from $T_c$ to 300 K confirming that the sum rule is satisfied \cite{tinkham1959}.
However, with lowering cutoff frequency $\omega_c$, the sum rule is being deviated, as it should be, but in a non-trivial way.
First, the normalized "partial" spectral weight $SW(T; \omega_c)/SW(300$ $K; \omega_c)$ with  $1,000$ $cm^{-1} < \omega_c < 10,000$ $cm^{-1}$ is monotonically decreasing with lowering temperature down to $T_c$. This means that the spectral weight below the cutoff frequency $\omega_c$ is transferred to the higher frequency region above the cutoff frequency $\omega_c$ when temperature decreases, which is an opposite behavior from a standard Drude type metallic state. Second, the rate of this spectral weight transfer is not monotonously increasing with lowering the cutoff frequency $\omega_c$; the deceasing rate increases to the maximum when the cutoff frequency is lowered to $\omega_c = 4,000 cm^{-1}$ (blue triangles) and then it becomes weaker with lowering cutoff frequencies to $\omega_c = 2,000$ $cm^{-1},$ and $1,000$ $cm^{-1}$. Third, in particular, when $\omega_c = 1,000$ $cm^{-1}$, the partial sum rule is "almost" recovered again: namely, the spectral weight below $\omega_c^{\ast} = 1,000$ $cm^{-1}$ (red circles) is separately conserved with respect to the temperature variation up to $T_c$.

Fig.4(B), on the other hand, shows the similar plots of $SW(T; \omega_c)/SW(300$ $K; \omega_c)$ with $\omega_c \leq \omega_c^{\ast} (=1,000 cm^{-1})$. It clearly shows that the partial spectral weight is "increasing" -- not decreasing --with decreasing temperature. This is an opposite behavior to the ones in Fig.4(A).

This complicated spectral weight transfer of $SW(T; \omega_c)$ with temperature variation is summarized in Fig.5. It shows that the spectral weight transfer of $\sigma_1(\omega)$ with temperature variation is roughly divided into two parts: one part below $\omega_c^{\ast} = 1,000$ $cm^{-1}$ and the other part above $\omega_c^{\ast} = 1,000$ $cm^{-1}$, and each part separately conserves the spectral weight. The physical meaning of it is that the correlated electrons are divided into the low energy itinerant part (Drude spectra) and the high energy localized part (Lorentzian spectra). At high temperatures, a large portion of spectra in the intermediate energy range -- in this particular compound, in between 500 $cm^{-1}$ and 5000 $cm^{-1}$ --  is undetermined, hence remains incoherent. Lowering temperature, these incoherent spectra are continuously depleted and transferred either into the low frequency Drude part or into the high frequency Lorentzian part: the final fate of the incoherent spectra was roughly predetermined by the frequency $\omega_c^{\ast} = 1,000$ $cm^{-1}$ -- which is the deepest valley point of the $\sigma_1(\omega)$ spectra in Fig.5 -- in this particular 10-3-8 compound. The sharpening of the Drude spectra with decreasing temperature is well understood as a development of coherent quasiparticles. The spectra transfer of the high frequency localized part should be associated with the strong correlation effects of the local interactions such as Hubbard interactions ($U, U^{'}$) and Hund's couplings ($J$) \cite{wang2012,schafgans2012,xu2014}.

Finally, in order to scrutinize the development of hump structure around 80-150 $cm^{-1}$ with temperature variation, we analyzed the low frequency spectra transfer in Fig.4(C),  with finer variation of cutoff frequencies for $\omega_c \leq 200  cm^{-1}$: 200, 100, 80, 60, and 40 $cm^{-1}$, respectively. The main observation is that there is a sudden change of the slopes of the spectral weight transfer with the cutoff frequency $\omega_c$ below and above $\omega_c = 100$ $cm^{-1}$; the slopes suddenly increase when $\omega_c \leq 100$ $cm^{-1}$. This means that decreasing temperature below $\sim 130 K$, there appears another drain of spectral weight transfer to the region in between $100 cm^{-1} - 200 cm^{-1}$ besides the narrowing of Drude spectra. This another drain of spectral weight transfer is the formation of "hump" structure around $80 cm^{-1}-150 cm^{-1}$ as seen with the 40K and 80K data in Fig.3.

\subsection{Drude-Lorentz model fitting for $\sigma_1^{exp}(\omega; T> T_c)$ in the normal state}

We have shown that the total spectra of the normal state optical conductivity $\sigma_1(\omega)$ consists of two spectral parts --  Drude part and Lorentz part -- and each part conserves their spectral weights: this separate conservation of sum rule will be confirmed once more in this section. Therefore we tried to fit the data with the standard Drude-Lorentz model
\begin{equation}
\sigma_1(\omega) = \frac{1}{4 \pi} Re \Big[ \sum_j \frac{\omega^2 _{p,j}}{\frac{1}{\tau_{D,j}}-i \omega} + \sum_k S_k \frac{\omega}{\frac{\omega}{\tau_{L,k}}  + i (\omega^2_{L,k} -\omega^2)} \Big]
\end{equation}
\noindent
where $\omega_{p,j}$ and $\frac{1}{\tau_{D,j}}$ are the plasma frequency and the scattering rate for the $j$-th free carrier Drude band, respectively,
and $S_k$, $\omega_{L,k}$  and  $\frac{1}{\tau_{L,k}}$  are the spectral weight, the Lorentz oscillator frequency, and the scattering rate of the $k$-th oscillator, respectively.

\begin{figure}[h]
\noindent
\includegraphics[width=90mm]{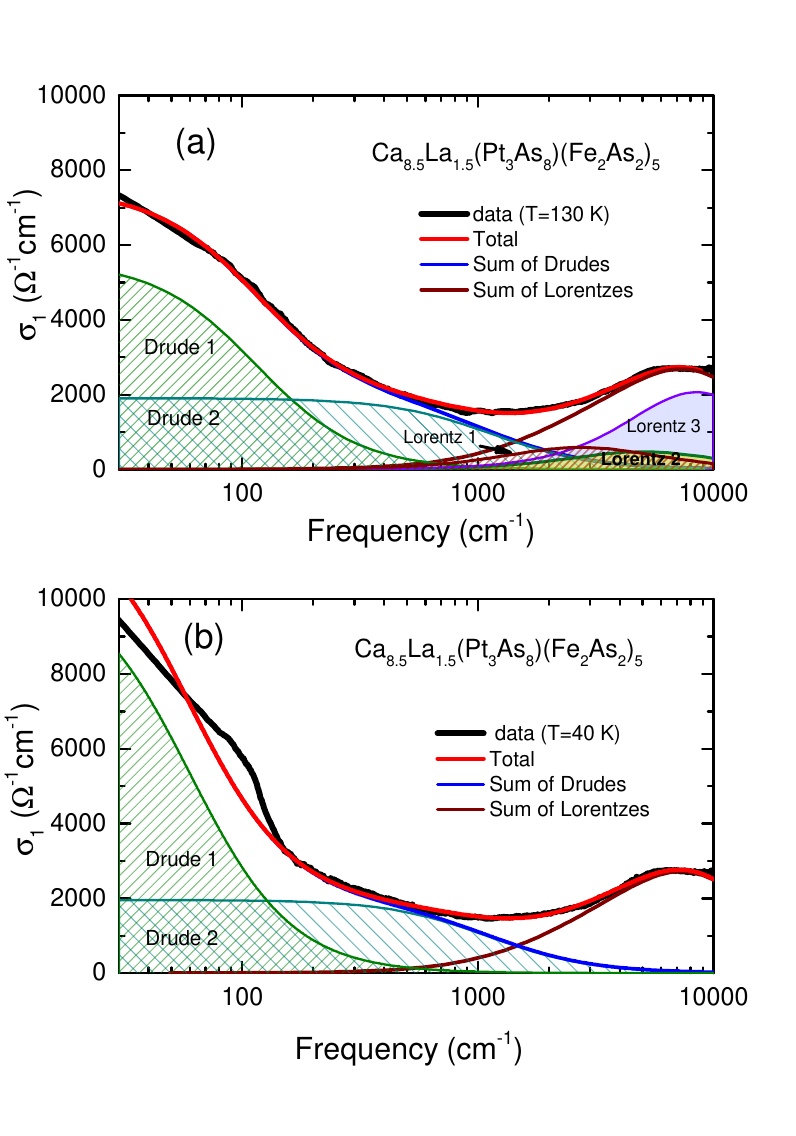}
\vspace{-1.cm}
\caption{(Color online) Typical Drude-Lorentz model fittings of the optical conductivity $\sigma_1(\omega)$ in the normal state: (A) 130 K data; (B) 40 K data.
Both data were fitted with two Drude forms and three Lorentz oscillators. The Lorentz oscillator parts and Drude-2 form do not change much with temperatures, but the Drude-1 form becomes substantially sharpened with decreasing temperature. \label{fig6}}
\end{figure}

\begin{figure}
\noindent
\includegraphics[width=90mm]{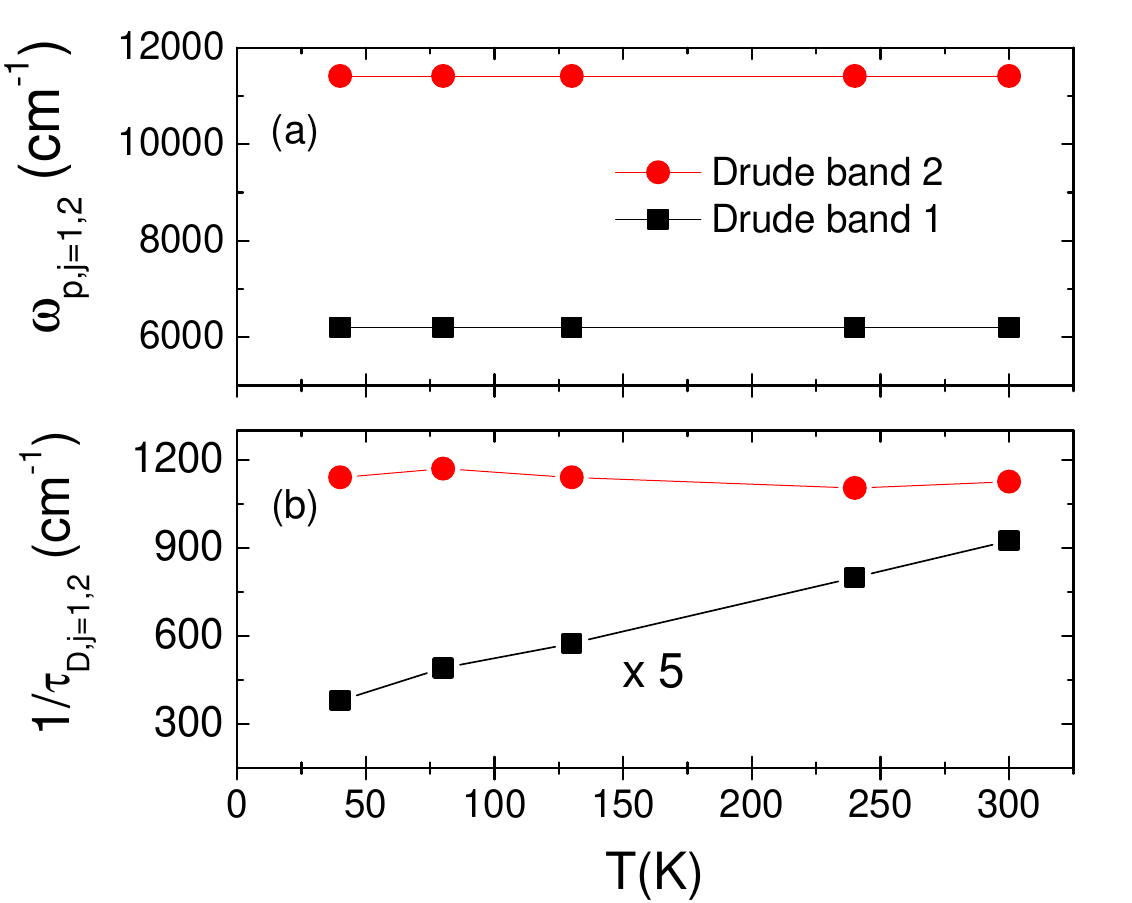}
\vspace{-0.5cm}
\caption{(Color online) (a) The plasma frequency of the Drude bands,  $\omega_{p1}$ (black solid square), and $\omega_{p2}$ (red solid circle). (b) The corresponding scattering rates, $1/\tau_{D1}$ (black solid square), and $1/\tau_{D2}$ (red solid circle). \label{fig7}}
\end{figure}

Figure 6(A) shows that the 130 K  data of $\sigma_1(\omega)$ and its Drude-Lorentz model fitting.
It is well fitted with two Drude terms (one narrow and the other broad) and three Lorentz oscillator terms.
Here two Drude terms show the presence of multiple bands in Ca$_{8.5}$La$_{1.5}$(Pt$_3$As$_8$)(Fe$_{2}$As$_{2}$)$_5$, as in many IBS compounds, e.g., LiFeAs compound\cite{min2013}.
The data of $\sigma_1(\omega)$ of 300 K and 240 K are also well fitted with the Drude-Lorentz model with the similar parameters:
the spectral weights, $\omega^2_{p,j=1,2}$ and $S_{k=1,2,3}$, and the Lorentz oscillator frequencies $\omega_{L,k=1,2,3}$ are the same,
but only the scattering rates, $\frac{1}{\tau_{D,j=1,2}}$ and $\frac{1}{\tau_{L,k=1,2,3}}$, need to be adjusted (see Fig.7).

However, the 40 K and 80 K data are different.
While the Lorentzian part can be fitted as above only with the scattering rates $\frac{1}{\tau_{L,k=1,2,3}}$ adjusted,
the low frequency Drude part has an extra hump structure around  $\sim 100$ $cm^{-1}$, as seen in Fig.3, hence can not be fitted by smooth Drude spectra.
Nevertheless, it was shown in the previous section that the low frequency spectra of $\sigma_1(\omega;T)$ for $\omega < 1,000$ $cm^{-1}$ separately conserve the spectral weight with temperature variation (see the pink inverse triangle data in Fig.4). Therefore, we tried to continue to use the same Drude-Lorentz model to
fit the 40 K and 80 K data.

Figure 6(B) shows the fitting result of the 40 K data. The Lorentzian part is fitted well as in the case of $T=130$ K, confirming the separate sum rule, but the Drude part show a clear deviation. Comparing the experimental data (black solid line) and the model fitting (red solid line), it shows that the hump structure (extra spectral peak) around $\omega \sim 100$ $cm^{-1}$ is formed by draining a spectral density from the lower frequency part ($\omega < 50$ $cm^{-1}$) of Drude spectra. Combining with the separate sum rule of the lower frequency $\sigma_1(\omega;T)$ for $\omega < 1,000$ $cm^{-1}$, this hump spectra should be due to a partial gapping in the Drude spectra.  If it were from an additional inter-band transition\cite{Heumen}, the sum rule should have been violated.
Impurity bound state scenario is also unrealistic because any potential scattering cannot form a bound state inside a Drude spectra without first forming a deep gap.
%

To summarize Fig.6, (1) the spectral weight of the hump structure around $\omega \sim 100$ $cm^{-1}$ is drained from the Drude spectra with a partial gapping in them; (2) this hump structure at 80 K and 40 K continuously evolve into the SC gap structure below $T_c$. We then conclude that the most plausible scenario for the hump structure around $\omega \sim 100$ $cm^{-1}$ is due to the preformed Cooper pairs.

\begin{table}[h]
\caption{Fitting parameters of Lorentz Oscillators} 
\centering 
\begin{tabular}{l c c c c} 
\hline\hline 
  & $\omega_{L,k=1,2,3}$ & T  & $S_k$  & $1/\tau_{L,k}$ \\
  & $(cm^{-1})$ & (K) & $(cm^{-1})$  & $(cm^{-1})$ \\ [0.5ex]
\hline 
 & & 40K &14224 & $5700$  \\[-1ex]
\raisebox{1.5ex}{Lorentz \#1} & \raisebox{1.5ex}{2734} & 130K
& 14224 & $5744$ \\[1ex]
& & 40K & 16811 & $8371$  \\[-1ex]
\raisebox{1.5ex}{Lorentz \#2} & \raisebox{1.5ex}{5000}& 130K
&16812 & $10000$  \\[1ex]
& & 40K & 42977 & $15000$  \\[-1ex]
\raisebox{1.5ex}{Lorentz \#3} & \raisebox{1.5ex}{8500}& 130K
&43319 & $15854$  \\[1ex]
\hline 
\end{tabular}
\label{tab:PPer}
\end{table}

In Fig.7, we show the temperature dependence of the fitting parameters
of the Drude spectra of the normal state optical conductivity $\sigma_1(\omega)$, i.e., plasma frequencies of the Drude bands,  $\omega_{p,i=1,2}$, and the corresponding scattering rates, $1/\tau_{D,i=1,2}$, for all measured temperatures. It is interesting to note the $T$-linear scattering rates $1/\tau_{D,1}(T)$ of the narrow Drude band in Fig.7(b) and the $T$-linear resistivity data $\rho_{DC}(T)$ in Fig.1, consistently indicating that the optimal doped Ca$_{8.5}$La$_{1.5}$(Pt$_3$As$_8$)(Fe$_{2}$As$_{2}$)$_5$ is located near the QCP, as discussed in the Introduction section. A similar behavior was also observed with the optimally doped Ba$_{0.6}$K$_{0.4}$Fe$_2$As$_2$ by optical spectroscopy\cite{Dai2013}, indicating the proximity to the QCP. However, the case of Ba$_{0.6}$K$_{0.4}$Fe$_2$As$_2$ is an antiferromagnetic QCP \cite{Dai2013}, while our case of Ca$_{8.5}$La$_{1.5}$(Pt$_3$As$_8$)(Fe$_{2}$As$_{2}$)$_5$ is a non-magnetic QCP, located far away from the AFM QCP\cite{ni2013}.
Finally, Table I shows the fitting parameters for the Lorentz oscillators of Fig.6: the 40K and 130K data.

\subsection{$\sigma_1^{exp}(\omega; T<T_c)$ in the SC state}

The three data set of $\sigma_1^{exp}(\omega; T<T_c)$ for $T=$8 K, 17 K, and 28 K, respectively, shown in Fig.3, clearly shows clean opening of a full gap, indicating an $s$-wave pairing. Combining with the presence of two Drude terms in the low frequency part of $\sigma_1^{exp}(\omega; T>T_c)$ in the normal state, as shown in the previous section, it is natural to assume two $s$-wave gaps in the SC state. Indeed, the line shape of the 8 K data (black solid line) of $\sigma_1^{exp}(\omega)$ appears to have two gaps with different sizes, $\Delta_1$ and $\Delta_2$.
Therefore, we use the generalized Mattis-Bardeen formula for two band superconductor to fit the 8 K data of $\sigma_1(\omega)$. The spectral weight of two band and its scattering rate  were taken as fitting parameters for optimal fitting. The results are shown in Fig.8.

\begin{figure}
\noindent
\includegraphics[width=100mm]{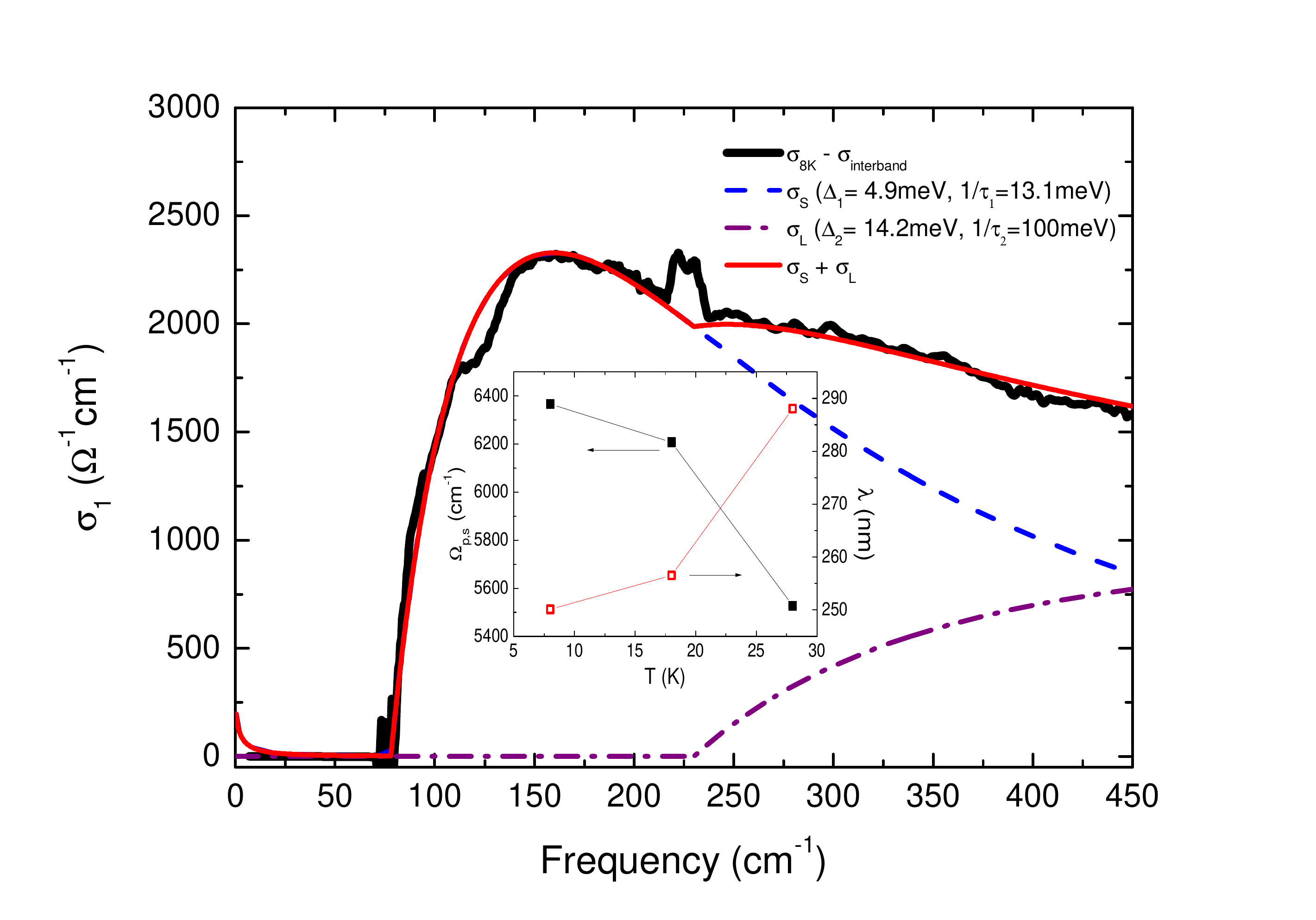}
\vspace{-0.cm}
\caption{(Color online) Mattis-Bardeen model fitting for the 8 K data (black solid line) of $\sigma_1^{exp}(\omega)$ after the Lorentz oscillator contribution subtracted. The data are decomposed into two Mattis-Bardeen terms (blue dashed line, and purple dash dotted line). The sum of the two Mattis-Bardeen terms is plotted as a red solid line. The inset represents the SC plasma frequency $\Omega_{p,S}(T)$ (black solid square) and the corresponding penetration depth $\lambda(T)$ (red open square). \label{fig8}}
\end{figure}

The black solid line in Fig. 8 shows the SC optical conductivity after subtracting the Lorentz oscillator contribution at high frequencies,
which is almost independent of temperature, from the optical conductivity of 8 K. The data shows the absorption edge at about 80 $cm^{-1}$, proportional to the first SC gap ($\sim 2\Delta_1$), and a weak kink at about 220 $cm^{-1}$ indicating a second SC gap ($\sim 2\Delta_2$).
With the generalized Mattis-Bardeen model \cite{zimmermann1991}, the SC gap sizes (the scattering rates) were determined to be $\Delta_1$ = 4.9 $meV$ ($1/\tau_1 = 13.1$ $meV$) and $\Delta_2$ = 14.2 $meV$ ($1/\tau_2 = 37.1$ $meV$), respectively.
The ratios of gap magnitude to $T_c$, $R=2\Delta/k_BT_c$, were evaluated as 3.6 and 10.2, respectively. These mixed values of $R$, compared to the value of $R_{BCS}=3.5$, suggest that the superconductivity in Ca$_{8.5}$La$_{1.5}$(Pt$_3$As$_8$)(Fe$_{2}$As$_{2}$)$_5$ could be a mixture of weak coupling and strong coupling SC states. To our knowledge, there is not yet an experimental report of the gap sizes of 10-3-8 compound. However, our gap values are consistent with the gap values of a typical FeAs-SC compounds with similar $T_c$; e.g., Ba$_{0.6}$K$_{0.4}$Fe$_2$As$_2$ ($T_c =37K$; $\Delta_1 \sim 6 meV, \Delta_2 \sim 12 meV$) \cite{ding2008}.

The relative sizes of the plasma frequency of each band are 1 (small gap band) to 2 (large gap band), consistent with the normal state Drude weights of two band. The scattering rate of the small gap band $1/\tau_1 = 13.1$ $meV \sim 104 cm^{-1}$ is slightly larger than the value of $1/\tau_{D1} \sim 75 cm^{-1}$ at the normal state, while the scattering rate of the larger gap band $1/\tau_2 = 100$ $meV \sim 800 cm^{-1}$ is smaller than the value of $1/\tau_{D2} \sim 1100 cm^{-1}$; the former is physically not very consistent while the latter is more reasonable. This inconsistency comes from the difference of two conductivity formula --  Mattis-Bardeen model \cite{zimmermann1991} and the Drude model --  which are not continuously connected. Considering this, the small inconsistency of the estimated scattering rates of the narrow band is in an understandable range.

In the inset, we also calculated the condensation strength  (the SC plasma frequency $\Omega_{p,S}(T)$) from the missing spectral weight and the corresponding penetration depth $\lambda(T)$ using the London formula $\Omega^2_{p,S}(T)=c^2/\lambda^2(T)$. Having only three data points below $T_c$ (8K, 17K, and 28K, respectively), we cannot extract much about the temperature dependence of these quantities. However, the overall trend  of temperature variation -- a flatter behavior at lower temperatures (i.e., $0< T < T_c /2$) and a rapid collapse near $T_c$ (i.e., $T_c /2 < T < T_c$) -- is very consistent with the temperature dependence for a s-wave SC gap(s). Also the absolute magnitude of the low temperature penetration depth $\lambda(T=8K) \sim 250 nm \sim 0.25 \mu m$ belongs to the standard range ($0.2\mu m < \lambda_{ab}(0) < 0.4\mu m$) of typical FeAs-superconductors \cite{prozorov2011}.

\subsection{$\sigma_1^{exp}(\omega; T_c <T \leq 80$ $K)$: Pseudo Gap and Preformed Cooper Pair Model}

\begin{figure}
\noindent
\includegraphics[width=100mm]{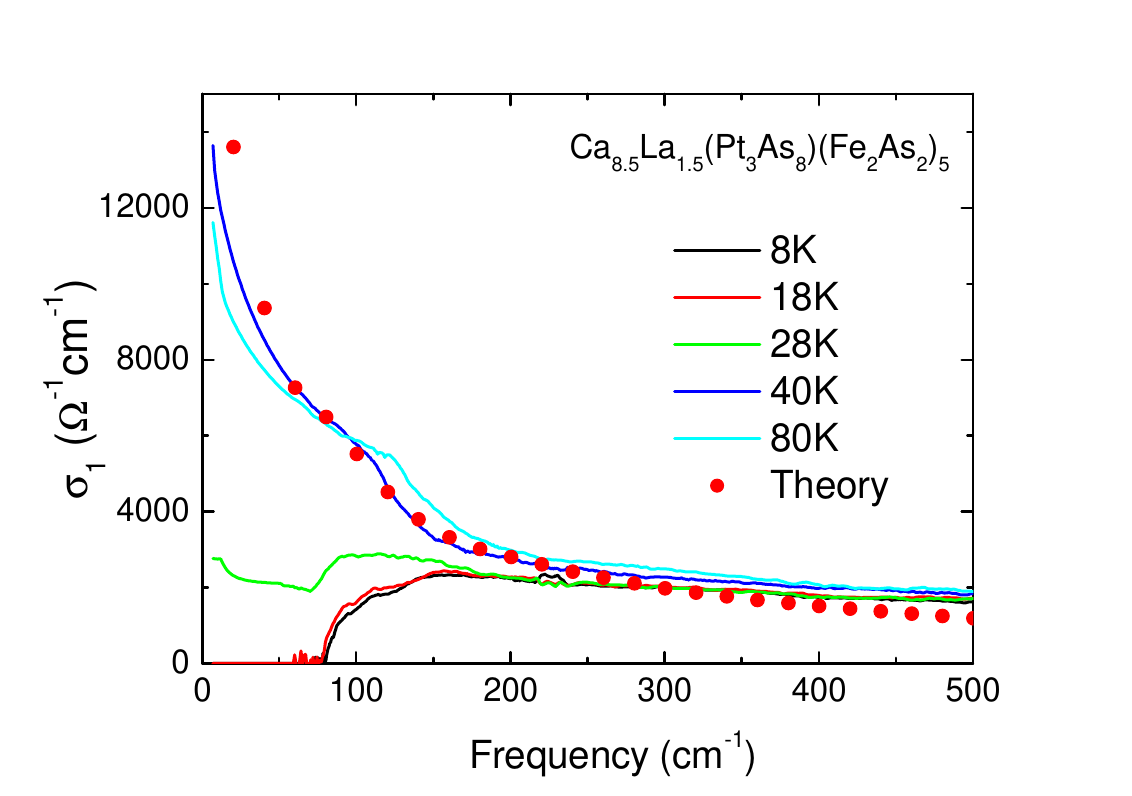}
\vspace{-0.cm}
\caption{(Color online) Theoretical calculation of $\sigma_1^{theo}(\omega, T > T_c)$ by the incoherent preformed Cooper pair model with $T=40$ K, overlayed on the experimental data of $\sigma_1^{exp}(\omega,T)$ for $T$ below and above $T_c=32.8$ K.
\label{fig9}}
\end{figure}

In order to understand the data of $\sigma_1^{exp}(\omega,T)$ above $T_c$ but below 80 K, we adopted the phase incoherent preformed Cooper pair model \cite{Emery95}.
In the ordinary SC state below $T_c$, the real part of the optical conductivity is calculated by a standard Kubo formula with sum of two bands $(a = 1, 2)$ as follows,
\beq
\sigma_1 (\omega,T)=-\frac{\sum_a Im \Lambda_{a,xx}
(\omega)}{\omega}
\eeq
with
\ba && Im\Lambda_{a,xx} (\omega)= \pi e^2 \int d^2 k d\omega^{'}
v_{a,x}^2 (k) \times \nonumber \\ & & Tr [\bar{A}_a
(k,\omega+\omega^{'} ) \bar{A}_a (k,\omega^{'} ) ]
[f(\omega+\omega^{'} )-f(\omega)] \ea
where $f(\omega)$ is the Fermi-Dirac distribution function, and $v_{a,x}$ is the Fermi velocity along the $x$-direction of the quasiparticles of the band "$a$". And $\bar{A}_a (k,\omega)$ is the $2 \times 2$ spectral density matrix of Nambu green's function of the band "$a$" in the SC state defined as $\bar{A}_a (k,\omega)=-Im \bar{G}_a (k,\omega)/\pi$ with
\beq \bar{G}_a (k,\omega)= \frac{\tilde{\omega}  \tau_0+\xi_a (k)
\tau_3 + \Delta_a \tau_1}{\tilde{\omega} ^2-\xi_a^2 (k)-\Delta_a^2
} \eeq
where $\Delta_a$ is the SC gap of the band "$a$" and $\tau_i$ are Pauli matrices.
The damping rate $\Gamma_a$ is included in $\tilde{\omega} = \omega + i\Gamma_a$ to fit the overall line shape, and without knowing $v_{a,x}$, the combined factor $v^2_{a,x} \int d^2k$ are taken as the parameter of the relative spectral weight of the each band $N^0_{a=1,2}$.

To simulate the phase fluctuations, we followed the recipe of Franz and Millis \cite{Franz_Millis98} and averaged the Nambu green's function Eq.(5) with Doppler shift energy
$\eta$ in the quasiparticle excitations as
\beq \tilde{\bar{G}}(k,\omega)= \int d \eta P(\eta) \bar{G}
(k,\omega-\eta) \eeq
where the probability distribution of $\eta$ is given by $P(\eta) = \sqrt{2 \pi W} e^{-\eta^2/2W}$ with $W \approx 3.48 \alpha_v
(T/T_c)\Delta^2$. $\alpha_v$ is a parameter derived from the XY-model and was estimated $\approx 0.009$ in the high-$T_c$
cuprates by Franz and Millis \cite{Franz_Millis98}, for example.

Here, we take the whole $W$ as a fitting parameter and chose $W_a = 0.06\Delta_{a} ^2$, which corresponds to $T \approx 1.4 T_c$, for the best fitting of the 40 K optical conductivity data. The results are plotted in Fig.9 by open red circle symbols.
We used the same gap values of $\Delta_1=4.9$ $meV$ and $\Delta_2=14.2$ $meV$ of Mattis-Bardeen model fitting in the SC state for 8 K data, and adjusted the scattering rates as $1/\tau_1=13$ $meV$ and $1/\tau_2=100$ $meV$, with cutoffs for $\omega < 2 \Delta_{a=1,2}$, respectively. The overall fitting is reasonably good and, in particular, it demonstrates that the hump structure around $100$ $cm^{-1}$ is qualitatively and quantitatively reproduced by the fluctuations of the incoherent SC OPs $\Delta_{a=1,2}$.

Finally, we have a remark on a possibility of a parasitical domain of undoped 10-3-8 phase as an origin for the hump structure above $T_c$ studied here. Although this kind of possibility of domain mixture always exists, we think this possibility is very low because: (1) the hump structure ($\sim 100 cm^{-1}$) has too small energy scale to be from the magnetic (SDW) ordering with $T_{Neel}=120K$, because, for example,  BaFe$_2$As$_2$ has a SDW ordering at 138K (a similar Neel temperature) but the optical data of  BaFe$_2$As$_2$ \cite{hu2010} shows that double humps appear at $360 cm^{-1}$ and $890 cm^{-1}$, respectively -- far higher than $\sim 100 cm^{-1}$ ;
(2) our hump structure at ($\sim 100 cm^{-1}$) continuously evolves from the 80K and 40K data ($>T_c$) to the 28K, 17K, and 8K data ($<T_c$) of the SC gap structure. For this hump structure to be from a magnetic gap, we need a unusual accidental coincidence to have the almost same gap energy scale both for the magnetic gap and the SC gap.

\section{Conclusions}
We have measured the optical reflectivity $R(\omega, T)$  of the optimally doped Ca$_{8.5}$La$_{1.5}$(Pt$_3$As$_8$)(Fe$_{2}$As$_{2}$)$_5$ single crystal ($T_c$ = 32.8 K) for frequencies from 40 cm$^{-1}$ to 12,000 cm$^{-1}$  and for temperatures from 8 K to 300 K.
In the normal state for $T =$ 130 K, 240 K, and 300 K, the optical conductivity data $\sigma_1(\omega, T)$ are well fitted by the Drude-Lorentz model with two Drude forms and three Lorentz oscillators. We have also found that (1) despite a huge variation of the spectral weight redistribution with temperature variation, the $f$-sum rule is satisfied; (2) also, divided by $\omega^{\ast} \sim 1,000$ $cm^{-1}$, the spectral weights above and below $\omega^{\ast}$ are separately conserved, which suggests that the original bare conduction electrons are split into the low energy itinerant carriers (Drude spectra) and the high energy localized carriers (Lorentzian oscillators) due to the strong correlation effects.

In the SC state for $T=$ 28 K, 17 K, and 8 K, the $\sigma_1(\omega, T < T_c)$ data show a clean gap opening at $\sim 80$ $cm^{-1}$  and the second gap structure at $\sim 220$ $cm^{-1}$. The $\sigma_1(\omega, T=8$ $K)$ data is well fitted with two SC gaps, $\Delta_{1}= 4.9$ $meV$ and $\Delta_{2}= 14.2$ $meV$, consistent with the presence of two Drude bands observed in the normal state data.
From the estimated fraction of the condensate spectral weight below $T_c$, we have estimated that only about $\sim 29 \%$ \cite{note1} of the conduction band carriers participates in the SC condensation indicating that Ca$_{8.5}$La$_{1.5}$(Pt$_3$As$_8$)(Fe$_{2}$As$_{2}$)$_5$ is in dirty limit.

Most interestingly, the $\sigma_1(\omega, T=40, 80$ $K)$ data, in the intermediate temperature region above $T_c$ but below $\sim 80$ K, showed a PG-like hump structure at the exactly same energy scale as the SC gap energy on top of the smooth Drude-like spectra. We have demonstrated that this PG-like hump structure can be consistently fitted with the preformed Cooper pair model using the same SC gap values of $\Delta_{1,2}$ of the SC state. Our work showed that the optimally doped Ca$_{8.5}$La$_{1.5}$(Pt$_3$As$_8$)(Fe$_{2}$As$_{2}$)$_5$ is a multigap superconductor having a clear PG formed by incoherent preformed Cooper pairs up to 80 K (about three times of $T_c$), therefore added one more case to the list of the IBSs with the PG due to the SC correlation\cite{kwon2012,dai2012}.

\section*{Acknowledgments}
YSK was supported by the NRF grant 2015M2B2A9028507 and 2016R1A2B4012672. YKB was supported by NRF Grant 2016-R1A2B4-008758.

\appendix
%
\numberwithin{equation}{section}
\numberwithin{figure}{section}

\section{DC limit of conductivity $\sigma_1(T, \omega)$}

In Fig.A.1, we show the DC limits of $\sigma_1(T, \omega)$ in the normal state for 40, 80, 130, 240, and 300 K, respectively, overlayed with the DC conductivities directly obtained from the $\rho_{DC}(T)$ data of Fig.1 as $\sigma_{DC}(T)=1/\rho_{DC}(T)$. The agreement at the DC limit is excellent for all temperatures which demonstrates the quality of our reflectivity data $R(T, \omega)$ (Fig.2) and the faithfulness of the KK-transformation.

For the low frequency region below $60 cm^{-1}$, we use the Hagen-Rubens extrapolation formula $R_{HR}(T, \omega) = 1 -2 \sqrt{\frac{2 \epsilon_0 \omega}{\sigma_{DC}}}$,
where the only free parameter $\sigma_{DC}(T)$ can be substituted by the DC resistivity data of Fig.1 as $\sigma_{DC}(T)=1/\rho_{DC}(T)$. However, to make the best smooth connection between
$R_{HR}(T, \omega)$ and our measured data $R(T, \omega)$, in the region of $40 -60 cm^{-1}$ (our data exist from $40 cm^{-1}$), we allow some adjustment of the value $\sigma_{DC}$ in the Hagen-Rubens formula. The necessary adjustments were less than $10 \%$ for all temperatures as demonstrated in Fig.A.1, which shows the excellent agreement between $\sigma_1(T, \omega \rightarrow 0)$ and $\sigma_{DC}(T)=1/\rho_{DC}(T)$.

\begin{figure}
\noindent
\includegraphics[width=90mm]{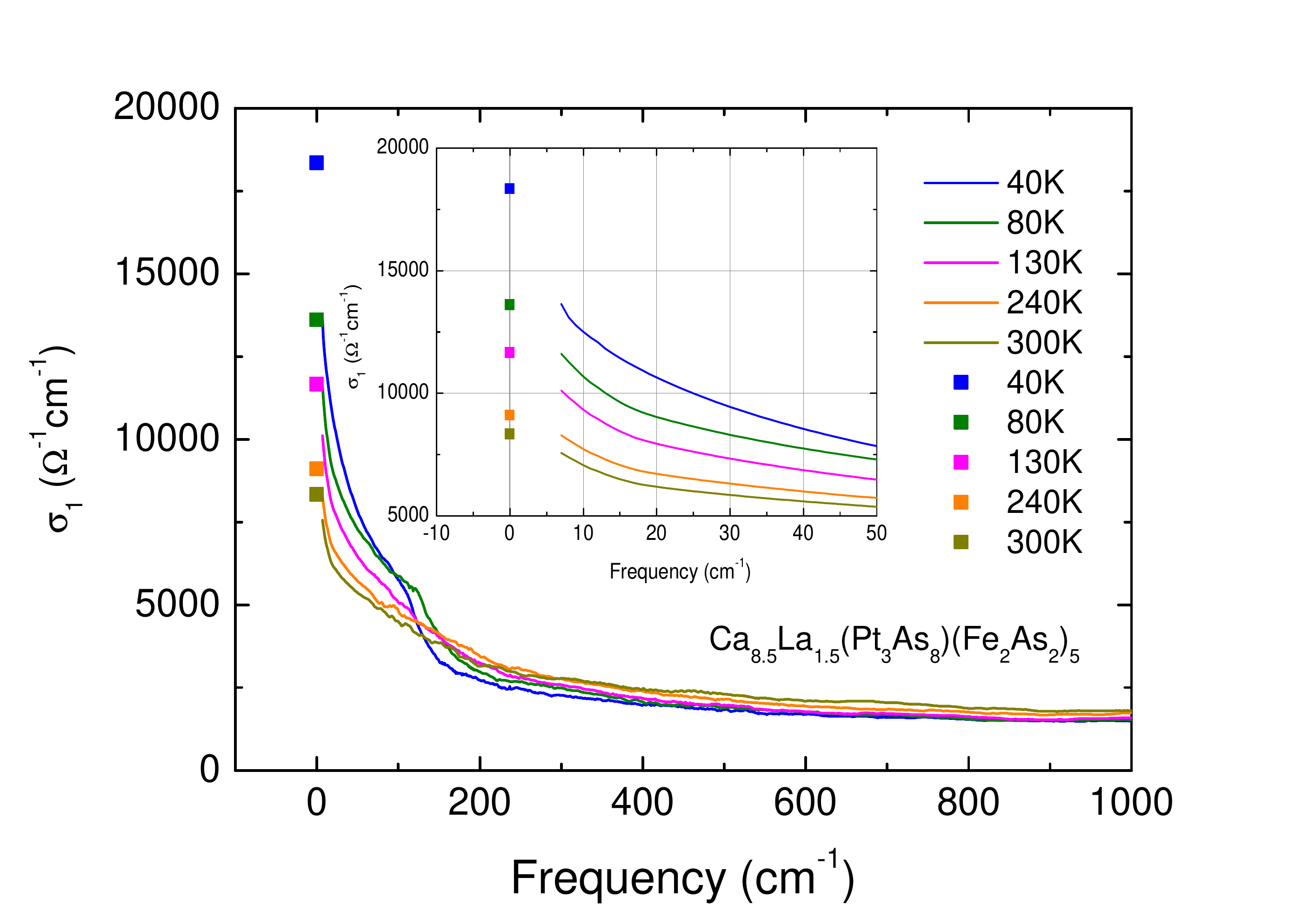}
\caption{(Color online) The same plot of Fig.3, $\sigma_1(\omega)$, but for wider $y$-axis range, to show the DC limit of $\sigma_1(\omega)$. The data are shown only for the normal state for 40, 80, 130, 240, and 300 K, respectively, and overlayed with the DC conductivities (solid square symbols) directly obtained from the data of Fig.1 as $\sigma_{DC}(T)=1/\rho_{DC}(T)$.    \label{FigA1}}
\end{figure}



\begin{references}
\bibitem{note1}
This condensate spectral weight is calculated from the total Drude spectral weight of the 40K data (Fig.6(B)) and the missing spectral weight of the 8K data (Fig.8).
This number $29 \%$ is smaller than the number $\sim 40 \%$ which can be read with the data (red circles) with the cutoff frequency $1000 cm^{-1}$ in Fig.4(b). This is because the Drude spectra have long tail beyond $\omega > 1000 cm^{-1}$.
\end{references}
\end{document}